\documentclass[12pt]{article}

\usepackage{amsmath, amsthm, amsfonts, amssymb}
\newtheorem{thm}{THEOREM}[section]
\newtheorem{cor}[thm]{COROLLARY}
\newtheorem{lem}[thm]{LEMMA}

\newtheorem{defn}[thm]{DEFINITION}

\newcommand{\dem}{{\it Proof.\ }}
\newcommand{\demthm}{{\it Proof of Theorem \ref{1}. }}
\newcommand{\fin}{$\square$}

\newcommand{\semifk}{$\left\{P^V_t\right\}_{t\geq 0}$}

\newcommand{\semi}{$\left\{T(t)\right\}_{t\geq 0}\;\;$}
\newcommand{\semia}{$\left\{T^{*}(t)\right\}_{t\geq 0}\;\;$}

\title{\bf Domains of uniqueness for $C_0$-semigroups on the dual of a Banach space}
\author{\sc Ludovic Dan Lemle}
\date{version 23 May 2008}

\def\D{\cal{D}}
\def\E{\mathbb{E}}

\def\N{\mathbb{N}}
\def\P{\mathbb{P}}

\def\R{\mathbb{R}}

\def\X{\cal{X}}

\begin{document}

\maketitle 


\noindent
{\footnotesize{\bf Abstract.}
 Let $({\cal X},\|\:.\:\|)$ be a Banach space. In general, for a $C_0$-semigroup \semi on $({\cal X},\|\:.\:\|)$, its adjoint semigroup \semia is no longer strongly continuous on the dual space $({\cal X}^{*},\|\:.\:\|^{*})$. Consider on ${\cal X}^{*}$ the topology of uniform convergence on compact subsets of $({\cal X},\|\:.\:\|)$ denoted by ${\cal C}({\cal X}^{*},{\cal X})$, for which the usual semigroups in literature becomes $C_0$-semigroups.\\ 
The main purpose of this paper is to prove that only a core can be the domain of uniqueness for a $C_0$-semigroup on $({\cal X}^{*},{\cal C}({\cal X}^{*},{\cal X}))$. As application, we show that the generalized Schr\"odinger operator ${\cal A}^Vf=\frac{1}{2}\Delta f+b\cdot\nabla f-Vf$, $f\in C_0^\infty(\R^d)$, is $L^\infty\left(\R^d,dx\right)$-unique. Moreover, we prove the $L^1(\R^d,dx)$-uniqueness of weak solution for the Fokker-Planck equation associated with ${\cal A}^V$.}

\section{Preliminaries}

A complete information on the general theory of strongly continuous semigroups of linear operators can be obtained by consulting the books of {\sc Yosida} \cite{yosida'71}, {\sc Davies} \cite{davies'80}, {\sc Pazy} \cite{pazy'83} or {\sc Goldstein} \cite{goldstein'85}.\\ 
In general, for a $C_0$-semigroup \semi on a Banach space $({\cal X},\|\:.\:\|)$, it is well known that its adjoint semigroup \semia is  no longer strongly continuous on the dual space $({\cal X}^{*},\|\:.\:\|^{*})$ with respect to the strong topology of ${\cal X}^{*}$. Without that strong continuity, the theory of semigroups becomes quite complicated and the Hille-Yosida theorem becomes very difficult (see {\sc Feller} \cite{feller'52}, \cite{feller'53}, {\sc Dynkin} \cite{dynkin'65}, {\sc Jefferies} \cite{jefferies'86}, \cite{jefferies'87} or {\sc Cerrai} \cite{cerrai'94}).\\
Recentely {\sc Wu} and {\sc  Zhang} \cite{wu-zhang'06} introduced on ${\cal X}^{*}$ a topology for which the usual semigroups in literature becomes $C_0$-semigroups. That is {\it the topology of uniform convergence on compact subsets of $({\cal  X},\|\:.\:\|)$}, denoted by ${\cal C}({\cal X}^{*},{\cal X})$.\\
It is not difficult to prove (see \cite[Lemma 1.10, p. 567]{wu-zhang'06})
\begin{lem}\label{28}
Let $({\cal X},\|\:.\:\|)$ be a Banach space. Then $({\cal X}^{*}, {\cal C}({\cal X}^{*},{\cal X}))$ is a locally convex space and:\\ 
i) the dual space $({\cal X}^{*}, {\cal C}({\cal X}^{*},{\cal X}))^{*}$ of $({\cal X}^{*}, {\cal C}({\cal X}^{*},{\cal X}))$ is $\cal X$;\\
ii) any bounded subset of $({\cal X}^{*}, {\cal C}({\cal X}^{*},{\cal X}))$ is $\|\:.\:\|^{*}$-bounded. And restriction to a $\|\:.\:\|^{*}$-bounded subset of $({\cal X}^{*}$, ${\cal C}({\cal X}^{*},{\cal X}))$ coincides with $\sigma({\cal X}^{*},{\cal X})$;\\ 
iii) $({\cal X}^{*}, {\cal C}({\cal X}^{*},{\cal X}))$ is complete;\\
iv) the topology ${\cal C}({\cal X},{\cal X}^{*}_{\cal C})$, where ${\cal X}^{*}_{\cal C}=\left({\cal X}^{*},{\cal C}({\cal X}^{*},{\cal X})\right)$, coincides with the $\|\:.\:\|$-topology of $\cal X$. 
\end{lem}
Moreover, if \semi is a $C_0$-semigroup on $({\cal X},\|\:.\:\|)$ with generator $\cal L$, then \semia is a $C_0$-semigroup on $({\cal X}^{*},{\cal C}({\cal X}^{*},{\cal X}))$ with generator ${\cal L}^{*}$ (see \cite[Theorem 1.4, p.564]{wu-zhang'06}). This is a satisfactory variant of Phillips theorem concerning the adjoint of a $C_0$-semigroup.\\
Therefore we have all ingredients to consider $C_0$-semigroups on the locally convex space $({\cal X}^{*}, {\cal C}({\cal X}^{*},{\cal X}))$. In accord to \cite[Definiton, p.234]{yosida'71}, we say that a family \semi of linear continuous operators on $({\cal X}^{*}, {\cal C}({\cal X}^{*},{\cal X}))$ is a {\it $C_0$-semigroup} on $({\cal X}^{*}, {\cal C}({\cal X}^{*},{\cal X}))$ if the following
properties holds:\\
(i) $T(0)=I$;\\
(ii) $T(t+s)=T(t)T(s)$, for all $t,s\geq 0$;\\
(iii) $\lim_{t\searrow 0}T(t)x=x$, for all $x\in({\cal X}^{*}, {\cal C}({\cal X}^{*},{\cal X}))$;\\
(iv) there exist a number $\omega_0\in\R$ such that the family
$\left\{e^{-\omega_0 t}T(t)\right\}_{t\geq 0}$ is equicontinuous.\\
The {\it infinitesimal generator} of the $C_0$-semigroup \semi is a
linear operator $\cal L$ defined on the domain
$$
{\cal D}({\cal L})=\left\{x\in{\X}\:\left|\:\lim_{t\searrow 0}\frac{T(t)x-x}{t}\mbox{ exists in }({\cal X}^{*}, {\cal C}({\cal X}^{*},{\cal X}))\right.\right\}
$$
by
$$
{\cal L}x=\lim_{t\searrow 0}\frac{T(t)x-x}{t}\quad,\quad\forall
x\in{\cal D}({\cal L}).
$$
We can see that $\cal L$ is a densely defined and closed operator on $({\cal X}^{*},{\cal C}({\cal X}^{*},{\cal X}))$ and the resolvent $R(\lambda;{\cal L})=(\lambda I-{\cal L})^{-1}$,
for any $\lambda\in\rho({\cal L})$ (the resolvent set of $\cal L$) satisfies the equality
$$
R(\lambda;{\cal L})x=\int\limits_{0}^{\infty}\!e^{-\lambda
t}T(t)x\:dt\quad,\quad\forall \lambda>\omega_0\mbox{ and }\forall
x\in{\cal X}^{*}.
$$
Unfortunately, in applications it is difficult to characterise completely the domain of generator ${\cal L}$. For this reason, sometimes we need to work on a subspace ${\cal D}\subset{\cal D}({\cal L})$ dense in $({\cal X}^{*}, {\cal C}({\cal X}^{*},{\cal X}))$ which is called a {\it core} of generator (see \cite[p.7]{davies'80}). More precisely,
\begin{defn}
\em
We say that ${\cal D}\subset{\cal D}({\cal L})$ is a core of generator ${\cal L}$ if ${\cal D}$ is dense in ${\cal D}({\cal L})$ with respect to the graph topology ${\cal C}_{\cal L}({\cal X}^{*},{\cal X})$ of ${\cal L}$ induced by the topology ${\cal C}({\cal X}^{*},{\cal X})$.
\end{defn}

This paper is organized as follows: in the next section by using a Desch-Schappacher perturbation of generator we prove that only a core can be the domain of uniqueness for a $C_0$-semigroup on $({\cal X}^{*},{\cal C}({\cal X}^{*},{\cal X}))$. This property is well known in the case of $C_0$-semigroups on Banach spaces (see \cite[Theorem 1.33, p.46]{arendt'86}), but here we prove it for a $C_0$-semigroup on the dual of a Banach space. In a forthcoming paper \cite{lemle-wu'08} we extend this property to the must difficult case of the dual of a locally convex space.

The Section 3 is devoted to study the $L^\infty\left(\R^d,dx\right)$-uniqueness of generalized Schr\"odinger operator. Remark that the natural topology for studying this problem is the topology of uniform convergence on compacts subsets of $\left(L^1\left(\R^d,dx\right),\|\:.\:\|_1\right)$ which is denoted by ${\cal C}\left(L^\infty,L^1\right)$.

In the first main result of Section 3 we find neccesary and sufficient conditions to show that the one-dimensional operator ${\cal A}_1^Vf=a(x)f^{''}+b(x)f^{'}-V(x)f$, $f\in C_0^\infty(x_0,y_0)$, where $-\infty\leq x_0<y_0\leq\infty$, is $L^\infty(x_0,y_0)$-unique.\\
In the second important result, by comparison with the one-dimensional case, we prove that the multidimensional generalized Schr\"odinger operator ${\cal A}^Vf=\frac{1}{2}\Delta f+b\cdot\nabla f-Vf$, $f\in C_0^\infty(\R^d)$ (where $\cdot$ is the iner product in $\R^d$), is $L^\infty\left(\R^d,dx\right)$-unique with respect to the topology ${\cal C}\left(L^\infty,L^1\right)$. As consequence, is obtained the $L^1\left(\R^d,dx\right)$-uniqueness of weak solution for the Fokker-Planck equation associated with ${\cal A}^V$. This result was reported in the conference EQUADIFF2007 held on August 2007 at Vienna.

\section{Uniqueness of pre-generators on the dual of a Banach space}

One of the main results of this paper concern the uniqueness of pre-generators on the dual of a Banach space. Recall that a linear operator ${\cal A}:{{\cal D}}\longrightarrow{\cal X}^{*}$ with the domain $\D$ dense in $({\cal X}^{*}, {\cal C}({\cal X}^{*},{\cal X}))$ is said to be {\it a pre-generator} in $({\cal X}^{*}, {\cal C}({\cal X}^{*},{\cal X}))$, if there exists some $C_0$-semigroup on $({\cal X}^{*}, {\cal C}({\cal X}^{*},{\cal X}))$ such that its generator $\cal L$ extends $\cal A$.\\ 
The main results of this section is
\begin{thm}\label{1}
Let ${\cal A}:{{\cal D}}\longrightarrow{\cal X}^{*}$ be a linear operator with
domain $\D$ dense in $({\cal X}^{*}, {\cal C}({\cal X}^{*},{\cal X}))$. 
Suppose that there exists a $C_0$-semigroup \semi on $({\cal X}^{*},{\cal C}({\cal X}^{*},{\cal X}))$ such that its generator $\cal L$ extends $\cal A$ (i.e. $\cal A$ is a pre-generator).\\
If $\cal D$ is not a core of $\cal L$, then there exists an infinite number of
extensions of $\cal A$ which are generators.
\end{thm}
For the proof of Theorem \ref{1} we need to use some perturbation result. Perturbation theory has long been a very useful tool in the hand of the analyst and physicist. A very elegant brief introduction to one-parameter semigroups is given in the treatise of {\sc Kato} \cite{kato'84} where on can find all results on perturbation theory. The perturbation by bounded operators is due to {\sc Phillips} \cite{phillips'53} who also investigate permanence of smoothness properties by this kind of perturbation. The perturbation by continuous operators on the graph norm of the generator is due to {\sc Desch} and {\sc Schappacher} \cite{desch-schappacher'84}.\\
Next lemma (comunicated by professor Liming Wu), which presents a Desch-Schappacher perturbation result for $C_0$-semigroups on $({\cal X}^{*}, {\cal C}({\cal X}^{*},{\cal X}))$, play a key rolle in the proof of Theorem \ref{1}:
\begin{lem}\label{desch-schappacher}
Let $({\cal X},\|\:.\:\|)$ be a Banach space, ${\cal L}$ the generator of a $C_0$-semigroup \semi on $({\cal X}^{*},{\cal C}({\cal X}^{*},{\cal X}))$ and $C$ a linear operator on $({\cal X}^{*},{\cal C}({\cal X}^{*},{\cal X}))$ with domain ${\cal D}({C})\supset{\cal D}({\cal L})$.\\
(i) If $C$ is ${\cal C}({\cal X}^{*},{\cal X})$-continuous, then ${\cal L}+C$ with domain ${\cal D}({\cal L}+C)={\cal D}({\cal L})$ is the generator of some $C_0$-semigroup on $({\cal X}^{*},{\cal C}({\cal X}^{*},{\cal X}))$.\\
(ii) If $C:{\cal D}({\cal L})\rightarrow{\cal D}({\cal L})$ is continuous with respect to the graph topology of $\cal L$ induced by the topology ${\cal C}({\cal X}^{*},{\cal X})$, then ${\cal L}+C$ with domain ${\cal D}({\cal L}+C)={\cal D}({\cal L})$ is the generator of some $C_0$-semigroup on $({\cal X}^{*},{\cal C}({\cal X}^{*},{\cal X}))$.
\end{lem}
\dem 
(i) By the \cite[Theorem 1.4, p.564]{wu-zhang'06} and using Lemma \ref{28}, ${\cal L}^{*}$ is the generator of the $C_0$-semigroup \semia on $({\cal X},{\cal C}({\cal X},{\cal X}^{*}_{\cal C}))=({\cal X},\|\:.\:\|)$. Under the condition on $C$, by \cite[Lemma 1.12, p.568]{wu-zhang'06} it follows that the operator ${C}^{*}$ is bounded on $({\cal X},\|\:.\:\|)$. By a well known perturbation result (see \cite[Theorem 1, p.68]{davies'80}), we find that ${\cal L}^{*}+{C}^{*}=({\cal L}+C)^{*}$ is the generator of some $C_0$-semigroup on $({\cal X},\|\:.\:\|)$. By using again \cite[Theorem 1.4, p.564]{wu-zhang'06}, we obtain that $({\cal L}+C)^{**}$ is the generator of some $C_0$-semigroup on $({\cal X}^{*},{\cal C}({\cal X}^{*},{\cal X}))$. Moreover, ${\cal D}(({\cal L}+C)^{*})$ is dense in $({\cal X},\|\:.\:\|)$. Hence ${\cal D}(({\cal L}+C)^{*})$ is dense in $({\cal X},\sigma({\cal X},{\cal X}^{*}))$. Then by \cite[Theorem 7.1, p.155]{schaefer'71} it follows that
$$
({\cal L}+C)^{**}=\overline{({\cal L}+C)}^{\sigma({\cal X}^{*},{\cal X})}
$$
Since $C$ is ${\cal C}({\cal X}^{*},{\cal X})$-continuous, by \cite[Lemma 1.5, p.564]{wu-zhang'06} it follows that $C$ is $\sigma({\cal X}^{*},{\cal X})$-continuous hence $\sigma({\cal X}^{*},{\cal X})$-closed. Consequently 
$$
{\cal L}+C=\overline{({\cal L}+C)}^{\sigma({\cal X}^{*},{\cal X})}
$$
from where it follows that $({\cal L}+C)^{**}={\cal L}+C$. Hence ${\cal L}+C$ is the generator of some $C_0$-semigroup on $({\cal X}^{*},{\cal C}({\cal X}^{*},{\cal X}))$.\\ 
(ii) We will follows closely the proof of {\sc Arendt} \cite[Theorem 1.31, p.45]{arendt'86}. Remark that $C:{\cal D}({\cal L})\rightarrow{\cal D}({\cal L})$ is continuous with respect to the graph topology of $\cal L$ induced by the topology ${\cal C}({\cal X}^{*},{\cal X})$ if and only if for all $\lambda>\omega_0$ (where $\omega_0$ is the real constatnte in the definition of the $C_0$-semigroup \semi) the operator
$$
\tilde{C}:=(\lambda I-{\cal L})CR(\lambda;{\cal L})
$$
is continuous on ${\cal X}^{*}$ with respect to the topology ${\cal C}({\cal X}^{*},{\cal X})$. Consequently, by (i) we find that ${\cal L}+\tilde{C}$ is the generator of some $C_0$-semigroup on $({\cal X}^{*},{\cal C}({\cal X}^{*},{\cal X}))$.
We shall prove that ${\cal L}+\tilde{C}$ is similar to ${\cal L}+C$. Remark that $C$ is continuous with respect to the graph norm $\|\:.\:\|^{*}+\|{\cal L}.\:\|^{*}$. By the prove of \cite[Theorem 1.31, p.45]{arendt'86}, there exists some $\lambda>\omega_0$ such that the operators 
$$
U:=I-CR(\lambda;{\cal L})\quad\mbox{and}\quad U^{-1} 
$$
are bounded on $({\cal X}^{*},\|\:.\:\|^{*})$. Moreover
$$
U({\cal L}+\tilde{C})U^{-1}=U({\cal L}-\lambda I+\tilde{C})U^{-1}+\lambda I=
$$
$$
=U[{\cal L}-\lambda I+(\lambda I-{\cal L})CR(\lambda;{\cal L})]U^{-1}+\lambda I=
$$
$$
=U({\cal L}-\lambda I)[I-CR(\lambda;{\cal L})]U^{-1}+\lambda I=
$$
$$
=U({\cal L}-\lambda I)+\lambda I=[I-CR(\lambda;{\cal L})]({\cal L}-\lambda I)+\lambda I=
$$
$$
={\cal L}-\lambda I+C+\lambda I={\cal L}+C
$$
Now we have only to prove that $U$ and $U^{-1}$ are continuous with respect to the topology ${\cal C}({\cal X}^{*},{\cal X})$. Since $CR(\lambda;{\cal L})=R(\lambda;{\cal L})\tilde{C}$ is continuous with respect to the topology ${\cal C}({\cal X}^{*},{\cal X})$, it follows that $U=I-CR(\lambda;{\cal L})$ is continuous with respect to the topology ${\cal C}({\cal X}^{*},{\cal X})$. 
On the other hand, by \cite[Lemma 1.5, p.564]{wu-zhang'06}, $U^{*}$ and $[CR(\lambda;{\cal L})]^{*}$ are continuous on $({\cal X},\|\:.\:\|)$. By Phillips theorem \cite[Proposition 5.9, p.246]{komatsu'64}, $1\in\rho([CR(\lambda;{\cal L})]^{*})$ if and only if $1\in[CR(\lambda;{\cal L})]^{**}$ and
$$
[I-([CR(\lambda;{\cal L})]^{*})^{-1}]^{*}=(I-[CR(\lambda;{\cal L})]^{**})^{-1}
$$
But by \cite[Theorem 1.1, p.155]{schaefer'71} we have $[CR(\lambda;{\cal L})]^{**}=CR(\lambda;{\cal L})$ and the right hand side above becomes $U^{-1}$. Hence $U^{-1}$, being the dual of some bounded operator on $({\cal X},\|\:.\:\|)$, is continuous on $({\cal X}^{*},{\cal C}({\cal X}^{*},{\cal X}))$ by \cite[Lemma 1.5, p.564]{wu-zhang'06} and the proof of lemma is completed. \fin\\
Now we are able to give\\
\demthm We will follows closely the proof of {\sc Arendt} \cite[Theorem 1.33, p.46]{arendt'86}. Endow ${\cal D}({\cal L})$ with the graph topology ${\cal C}_{\cal
L}({\cal X}^{*},{\cal X})$ of ${\cal L}$ induced by the topology ${\cal C}({\cal X}^{*},{\cal X})$. If in contrary
$\D$ is not a core of $\cal L$, then $\D$ is not dense in ${\cal
D}({\cal L})$ with respect to the graph topology ${\cal C}_{\cal
L}({\cal X}^{*},{\cal X})$ of $\cal L$. By Hahn-Banach theorem there exist some non-zero
linear functional $\phi$ continuous on ${\cal D}({\cal L})$ with
respect to the graph topology ${\cal C}_{\cal
L}({\cal X}^{*},{\cal X})$ of $\cal L$ such that $\phi(x)=0$
for all $x\in{\D}$. Fix some $u\in{\cal D}({\cal L})$, $u\neq 0$, and
consider the linear operator
$$
C:{\cal D}({\cal L})\longrightarrow{\cal D}({\cal L})
$$
$$
Cx=\phi(x)u\quad,\quad\forall x\in{\cal D}({\cal L}).
$$
Then $C$ is continuous with respect to the graph topology ${\cal C}_{\cal
L}({\cal X}^{*},{\cal X})$ of $\cal L$ on ${\cal D}({\cal L})$. By (Desch-Schappacher perturbation) Lemma \ref{desch-schappacher} it follows that ${\cal L}+C$ is the generator of some $C_0$-semigroupe on $({\cal X}^{*},{\cal C}({\cal X}^{*},{\cal X}))$ and 
$$
({\cal L}+C)/_{\cal D}={\cal L}/_{\cal D}={\cal A}
$$
It is obvious that an infinite number of generators can be constructed in that way. \fin

\section{$L^\infty(\R^d,dx)$-uniqueness of generalized Schr\"odinger operators}

In this section we consider the generalized Schr\"odinger operator
$$
{\cal A}^Vf:=\frac{1}{2}\Delta f+b\cdot\nabla f-Vf\quad,\quad\forall f\in C_0^\infty(\R^d)
$$
where $b:\R^d\rightarrow\R^d$ is a measurable locally bounded vector field and $V:\R^d\rightarrow\R$ is a locally bounded potential. The study of this operator has attracted much attention both from the people working on Nelson's stochastic mechanics ({\sc Carmona} \cite{carmona'85}, {\sc Meyer} and {\sc Zheng} \cite{meyer-zheng'84}, etc.) and from those working on the theory of Dirichlet forms ({\sc Albeverio}, {\sc Brasche} and {\sc R\"ockner} \cite{albeverio-brasche-rockner'89}). In the case where $V=0$, the essential self-adjointness of ${\cal A}:=\frac{1}{2}\Delta+b\cdot\nabla$ in $L^2$ has been completely charaterized in the works of {\sc Wielens} \cite{wielens'85} and {\sc Liskevitch} \cite{liskevitch'99}. $L^1$-uniqueness of this operator has been introduced and studied by {\sc Wu} \cite{wu'99}, its $L^p$-uniqueness has been studied by {\sc Eberle} \cite{eberle'00} for $p\in[1,\infty)$ and by {\sc Wu} and {\sc Zhang} \cite{wu-zhang'06} for $p=\infty$.

In accord with the Theorem \ref{1}, we can introduce $L^\infty\left(\R^d,dx\right)$-uniqueness of pre-generators in a very natural form:
\begin{defn}
\em
We say that a pre-generator $\cal A$
is $\left(L^\infty\left(\R^d,dx\right),{\cal C}\left(L^\infty,L^1\right)\right)$-unique, if there exists only one $C_0$-semigroup \semi on $\left(L^\infty\left(\R^d,dx\right),{\cal C}\left(L^\infty,L^1\right)\right)$ such that its generator $\cal L$ is an extension of $\cal A$. 
\end{defn}
This uniqueness notion has been used by {\sc Arendt} \cite{arendt'86}, {\sc
R\"ockner} \cite{rockner'98}, {\sc Wu} \cite{wu'98} and
\cite{wu'99}, {\sc
Eberle} \cite{eberle'00}, {\sc Arendt}, {\sc Metafune} and {\sc Pallara} \cite{arendt-metafune-pallara'06}, {\sc Wu} and {\sc Zhang} \cite{wu-zhang'06}, {\sc Lemle} \cite{lemle'07} and others in different contexts. The next characterisation of $\left(L^\infty\left(\R^d,dx\right),{\cal C}\left(L^\infty,L^1\right)\right)$-uniqueness of pre-generators is wery useful in applications (for others characterisations of the uniqueness of pre-generators we strongly recommanded for the reader the excelent article of {\sc Wu} and {\sc Zhang} \cite{wu-zhang'06}): 
\begin{thm}\label{11}
Let $\cal A$ be a linear operator on $\left(L^\infty\left(\R^d,dx\right),{\cal C}\left(L^\infty,L^1\right)\right)$ with domain $\D$ (the
test-function space) which is assumed to be dense in $\left(L^\infty\left(\R^d,dx\right),{\cal C}\left(L^\infty,L^1\right)\right)$. Assume that there is a $C_0$-semigroup \semi on $\left(L^\infty\left(\R^d,dx\right),{\cal C}\left(L^\infty,L^1\right)\right)$ such
that its generator $\cal L$ is an extension
of $\cal A$ (i.e., $\cal A$ is a pre-generator). The following assertions are equivalents:\\
(i) $\cal A$ is $\left(L^\infty\left(\R^d,dx\right),{\cal C}\left(L^\infty,L^1\right)\right)$-unique;\\
(ii) $\D$ is a core of $\cal L$;\\
(iii) for some $\lambda>\omega_0$ (where $\omega_0\in\R$ is the constant in
the definition of $C_0$-semigroup \semi),
the range $(\lambda I-{\cal A})(\D)$ is dense in $\left(L^\infty\left(\R^d,dx\right),{\cal C}\left(L^\infty,L^1\right)\right)$;\\
(iv) (Liouville property) for some $\lambda>\omega_0$, if $h\in{\cal D}({\cal A}^{*})$ satisfies $(\lambda I-{\cal A}^{*})h=0$, then $h=0$;\\
(v) (uniqueness of weak solutions for the dual Cauchy problem)
for every $f\in\left(L^1\left(\R^d,dx\right),\|\:.\:\|_1\right)$, the dual Cauchy problem
$$
\left\lbrace\begin{array}{l}
\partial_tu(t,x)={\cal A}^{*}u(t,x)\\
u(0,x)=f(x)
\end{array}
\right.
$$
has a $\left(L^1\left(\R^d,dx\right),\|\:.\:\|_1\right)$-unique weak solution $u(t,x)=T^{*}(t)f(x)$.
\end{thm}

Our main purpose in this section is to find some sufficient condition to assure the $L^\infty(\R^d,dx)$-uniqueness of $({\cal A}^V,C_0^\infty(\R^d))$ with respect to the topology ${\cal C}\left(L^\infty,L^1\right)$ in the case where $V\geq 0$.

At first, we must remark that the generalized Schr\"odinger operator $({\cal A}^V,C_0^\infty(\R^d))$ is a pre-generator on $\left(L^\infty(\R^d,dx),{\cal C}\left(L^\infty,L^1\right)\right)$. Indeed, if we consider the Feynman-Kac semigroup \semifk given by
$$
P_t^Vf(x):=\E^x1_{[t<\tau_e]}f(X_t)e^{-\int\limits_0^t\!V(X_s)\:ds}
$$
where $(X_t)_{0\leq t<\tau_e}$ is the diffusion generated by $\cal A$ and $\tau_e$ is the explosion time, then by \cite[Theorem 1.4]{wu-zhang'06} \semifk is a $C_0$-semigroup on $L^\infty(\R^d,dx)$ with respect to the topology ${\cal C}\left(L^\infty,L^1\right)$. Let $\partial$ be the point at infinity of $\R^d$. If we put $X_t=\partial$ after the explosion time $t\geq\tau_e$, then by Ito's formula it follows for any $f\in C_0^\infty(\R^d)$ that
$$
f(X_t)-f(x)-\int\limits_0^t\!{\cal A}^Vf(X_s)\:ds
$$
is a local martingale. As it is bounded over bounded times intervals, it is a true martingale. Thus by taking the expectation under $\P_x$, we get
$$
P_t^Vf(x)-f(x)=\int\limits_0^t\!P_s^V{\cal A}^Vf(x)\:ds\quad,\quad\forall t\geq 0.
$$
Therefore $f$ belongs to the domain of the generator ${\cal L}^V_{(\infty)}$ of $C_0$-semigroup \semifk on $(L^\infty(\R^d,dx),{\cal C}\left(L^\infty,L^1\right))$. Consequently, $({\cal A}^V,C_0^\infty(\R^d))$ is a pre-generator on $L^\infty(\R^d,dx)$ with respect to the topology ${\cal C}\left(L^\infty,L^1\right)$ and we can apply the Theorem \ref{11} to study the $(L^\infty(\R^d,dx),{\cal C}\left(L^\infty,L^1\right))$-uniqueness of this operator. 

\subsection{The one-dimensional case}
The purpose of this subsection is to study the $L^\infty$-uniqueness of one-dimensional operator
$$
{\cal A}_1^Vf=a(x)f^{''}+b(x)f^{'}-V(x)f\quad,\quad f\in C_0^\infty(x_0,y_0)
$$
where $-\infty\leq x_0<y_0\leq\infty$ and the coefficients $a$, $b$ and $V$ satisfy the next properties
$$
a(x),\:b(x)\in L_{loc}^\infty(x_0,y_0;dx)
$$
$$
V(x)\in L_{loc}^\infty(x_0,y_0;dx),\:\:V(x)\geq 0
$$
and the following very weak ellipticity condition
$$
a(x)>0\quad dx-\mbox{a.e.}
$$
$$
\frac{1}{a(x)},\quad\frac{b(x)}{a(x)}\in L_{loc}^1(x_0,y_0;dx)
$$
where $L_{loc}^\infty(x_0,y_0;dx)$ , respectively $L_{loc}^1(x_0,y_0;dx)$, denotes the space of real Lebesgue measurable functions which are essentially bounded, respectively integrable, with respect to Lebesgue measure on any compact sub-interval of $(x_0,y_0)$.\\
Fix a point $c\in(x_0,y_0)$ and let
$$
\rho(x)=\frac{1}{a(x)}e^{\int\limits_c^x\!\frac{b(t)}{a(t)}\:dt}\quad.
$$ 
be {\it the speed measure of Feller} and let
$$
\alpha(x)=e^{\int\limits_c^x\!\frac{b(t)}{a(t)}\:dt}
$$
be {\it the scale function of Feller}. It is easy to see that
$$
\left\langle {\cal A}_1^Vf,g\right\rangle_\rho=\left\langle f,{\cal A}_1^Vg\right\rangle_\rho\quad,\quad\forall f,g\in C_0^\infty(x_0,y_0)
$$
where
$$
\langle f,g\rangle_\rho=\int\limits_{x_0}^{y_0}\!f(x)g(x)\rho(x)\:dx\quad.
$$
For $f\in C_0^\infty(x_0,y_0)$, we can write ${\cal A}_1^V$ in the Feller form:
$$
{\cal A}_1^V=a(x)f^{''}+b(x)f^{'}-V(x)f=\frac{\alpha(x)}{\rho(x)}f^{''}+\frac{a(x)\alpha^{'}(x)}{\alpha(x)}f^{'}-V(x)f=
$$
$$
=\frac{\alpha(x)}{\rho(x)}f^{''}+\frac{\alpha^{'}(x)}{\rho(x)}f^{'}-V(x)f=\frac{1}{\rho(x)}\left[\alpha(x)f^{'}\right]^{'}-V(x)f
$$
and the assumptions concerning the coeficients $a(x)$ and $b(x)$ can be writen as
\begin{itemize}
	\item $\rho(x)>0$, $dx$-a.e. and $\rho\in L_{loc}^1(x_0,y_0;dx)$
	\item $\alpha(x)>0$ everywhere and $\alpha$ is absolutely continuous
	\item $\alpha/\rho,\quad\alpha^{'}/\rho\in L_{loc}^\infty(x_0,y_0;dx)$.
\end{itemize}
Now consider the operator $({\cal A}_1^V,C_0^\infty(x_0,y_0))$ as an operator on $L^\infty(x_0,y_0;\rho dx)$ which is endowed with the topology ${\cal C}(L^\infty(x_0,y_0,\rho dx),L^1(x_0,y_0,\rho dx))$. We begin with a series of lemmas.
\begin{lem}\label{21}
Let $({\cal A}_1^V)^{*}:{\cal D}(({\cal A}_1^V)^{*})\subset L^1(x_0,y_0;\rho dx)\rightarrow L^1(x_0,y_0;\rho dx)$ be the adjoint operator of ${\cal A}_1^V$. Let $\lambda>0$ and let $u\in L^1(x_0,y_0;\rho dx)$ be in ${\cal D}(({\cal A}_1^V)^{*})$ such that 
$$
({\cal A}_1^V)^{*}u=\lambda u.
$$ 
Then $u$ solves the ordinary differential equation
$$
\left(\alpha u^{'}\right)^{'}=\lambda u\rho+Vu\rho
$$
in the following sense: $u$ has an absolutely continuous $dx$-version $\hat{u}$ such that $\hat{u}^{'}$ is absolutely continuous and
$$
\left(\alpha\hat{u}^{'}\right)^{'}=\lambda\hat{u}\rho+V\hat{u}\rho.
$$
\end{lem}
\dem
The sufficiency follows easily by integration by parts.\\
Below we prove the necessity. Let $x_0<x_1<y_1<y_0$. The space of distributions on $(x_1,y_1)$ is denoted by ${\cal D}'(x_1,y_1)$.\\
{\bf (I)} We recall that if $k\geq 1$ and $T_1,T_2\in{\cal D}'(x_1,y_1)$ satisfy $T_1^{(k)}=T_2^{(k)}$ i.e. 
$$
\int\limits_{x_1}^{y_1}\!T_1f^{(k)}(x)\:dx=\int\limits_{x_1}^{y_1}\!T_2f^{(k)}(x)\:dx
$$
for any $f\in C_0^\infty(x_1,y_1)$, then there exists a polynomial $w$ such that $T_1=T_2+w$.\\
{\bf (II)} Let $u\in L^1(x_0,y_0;\rho dx)$ be in ${\cal D}(({\cal A}_1^V)^{*})$ such that
$$
({\cal A}_1^V)^{*}u=\lambda u.
$$
Then for $f\in C_0^\infty(x_1,y_1)$ we have:
\begin{eqnarray*}
& &\int\limits_{x_1}^{y_1}\!u\left(\alpha f^{'}\right)^{'}\:dx=\int\limits_{x_1}^{y_1}\!u{\cal A}_1^Vf\rho\:dx+\int\limits_{x_1}^{y_1}\!uVf\rho\:dx=\\
&=&\left\langle u,{\cal A}_1^Vf\right\rangle_\rho+\left\langle u,Vf\right\rangle_\rho=\left\langle ({\cal A}_1^V)^{*}u,f\right\rangle_\rho+\left\langle u,Vf\right\rangle_\rho=\\
&=&\left\langle \lambda u,f\right\rangle_\rho+\left\langle u,Vf\right\rangle_\rho=\lambda\int\limits_{x_1}^{y_1}\!uf\rho\:dx+\int\limits_{x_1}^{y_1}\!uVf\rho\:dx.
\end{eqnarray*}
From
$$
|f(x)|=\left|\int\limits_{x_1}^{x}\!f^{'}(t)\:dt\right|
\leq\int\limits_{x_1}^{x}\!|f^{'}(t)|\:dt\leq\int\limits_{x_1}^{y_1}\!|f^{'}(t)|\:dt
$$
it follows that
$$
\|f\|_{L^\infty(x_1,y_1;dx)}\leq\|f^{'}\|_{L^1(x_1,y_1;dx)}
$$
and we have
\begin{eqnarray*}
& &\left|\int\limits_{x_1}^{y_1}\!u\left[\alpha f^{''}+
\alpha^{'}f^{'}\right]\:dx\right|=\left|\int\limits_{x_1}^{y_1}\!u
\left(\alpha f^{'}\right)^{'}\:dx\right|\leq\\
&\leq&\lambda\left|\int\limits_{x_1}^{y_1}\!uf\rho\:dx\right|+\left|\int
\limits_{x_1}^{y_1}\!uVf\rho\:dx\right|\leq\\
&\leq&\left[\lambda\left\|u\rho\right\|_{L^1(x_0,y_0;dx)}+\left\|uV\rho\right
\|_{L^1(x_1,y_1;dx)}\right]\|f\|_{L^\infty(x_1,y_1;dx)}\leq\\
&\leq&C\left\|f^{'}\right\|_{L^1(x_1,y_1;dx)}
\end{eqnarray*}
where
$$
C=\lambda\left\|u\rho\right\|_{L^1(x_0,y_0;dx)}+\left\|uV\rho\right
\|_{L^1(x_1,y_1;dx)}
$$
is independent of $f$. The above inequality means that the linear functional
$$
l_u(\eta):=\int\limits_{x_1}^{y_1}\!u\left(\alpha\eta^{'}+
\alpha^{'}\eta\right)\:dx
$$
where $\eta\in\left\{f^{'}\:\left|\:f\in C_0^\infty(x_1,y_1)\right.\right\}\subset L^1(x_1,y_1;dx)$, is continuous with respect to the $L^1(x_1,y_1;dx)$-norm. Thus by the Hahn-Banach's theorem and the fact that the dual of $L^1(x_1,y_1;dx)$ is $L^\infty(x_1,y_1;dx)$, there exists $v\in L^\infty(x_1,y_1;dx)$ such that
$$
l_u(\eta):=\int\limits_{x_1}^{y_1}\!u\left(\alpha\eta^{'}+
\alpha^{'}\eta\right)\:dx=\int\limits_{x_1}^{y_1}\!v\eta\:dx
$$
which implies
$$
\int\limits_{x_1}^{y_1}\!u\alpha\eta^{'}\:dx=\int\limits_{x_1}^{y_1}\!
\left(v-u\alpha^{'}\right)\eta\:dx=\int\limits_{x_1}^{y_1}\!h\eta^{'}\:dx
$$
where
$$
h(x)=-\int\limits_{x_1}^{x}\!\left[v(t)-u(t)\alpha^{'}(t)\right]\:dt
$$
is an absolutely continuous function on $(x_1,y_1)$. It follows from {\bf (I)} that there exists a polynomial $w$ such that 
$$
u\alpha=h+w
$$
on $(x_1,y_1)$ in the sense of distributions, hence $u\alpha=h+w$ a.e. on $(x_1,y_1)$.\\
{\bf (III)} Since $\alpha>0$ is absolutely continuous, the equality 
$$
u=\alpha^{-1}(h+w)\quad\mbox{a.e.}
$$ 
shows that $u$ also has an absolutely continuous version 
$$
\tilde{u}:=\alpha^{-1}(h+w).
$$
{\bf (IV)} Now we have
\begin{eqnarray*}
\lambda\int\limits_{x_1}^{y_1}\!\tilde{u}f\rho\:dx&=&\int\limits_{x_1}^{y_1}\!\tilde{u}\left(
\alpha f^{'}\right)^{'}\:dx-\int\limits_{x_1}^{y_1}\!\tilde{u}Vf\rho\:dx=\\
&=&-\int\limits_{x_1}^{y_1}\!\tilde{u}^{'}\alpha f^{'}\:dx-\int\limits_{x_1}^{y_1}
\!\tilde{u}Vf\rho\:dx.
\end{eqnarray*}
so that 
$$
\int\limits_{x_1}^{y_1}\!\left(\lambda\tilde{u}\rho+\tilde{u}V\rho\right)\:dx=
-\int\limits_{x_1}^{y_1}\!\tilde{u}^{'}\alpha f^{'}\:dx.
$$
Hence
$$
\left(\alpha\tilde{u}^{'}\right)^{'}=\lambda\tilde{u}\rho+\tilde{u}V\rho\in L^1(x_1,y_1;dx)
$$
in the sense of distributions. Then $\alpha\tilde{u}^{'}$ has an absolutely continuous version, so is $\tilde{\tilde{u}}^{'}$ (a primitive of $\lambda\tilde{u}\rho+\tilde{u}V\rho$) on $(x_1,y_1)$ and
$$
\tilde{\tilde{u}}^{'}=\lambda\tilde{u}\rho+\tilde{u}V\rho\quad\mbox{a.e.}
$$
{\bf (V)} From the above discution we have
$$
\alpha\tilde{u}^{'}=\tilde{\tilde{u}}\quad\mbox{a.e.}
$$
which implies that
$$
\tilde{u}^{'}=\alpha^{-1}\tilde{\tilde{u}}\quad\mbox{a.e.}
$$
Since $\alpha^{-1}\tilde{\tilde{u}}$ is absolutely continuous, we get that $\tilde{u}$, hence $u$ has a version $\hat{u}$ (a primitive of $\alpha^{-1}\tilde{\tilde{u}}$) such that
$$
\hat{u}^{'}=\alpha^{-1}\tilde{\tilde{u}}
$$
is absolutely continuous. We then go back to {\bf (IV)}, using $\hat{u}$ in place of $\tilde{u}$, to obtain
$$
\left(\alpha\hat{u}^{'}\right)^{'}=\lambda\hat{u}\rho+V\hat{u}\rho.
$$
The lemma is thus proved since $(x_1,y_1)$ is an arbitrary relatively compact subinterval of $(x_0,y_0)$. \fin

\begin{lem}\label{22}
Let $\lambda>0$ and let $u\in L^1(x_0,y_0;\rho dx)$ be such that
$$
({\cal A}_1^V)^{*}u=\lambda u
$$
in the sense of Lemma \ref{21}. We may suppose that $u$ is an absolutely continuous version such that $u^{'}$ is absolutely continuous. Let $c_1\in(x_0,y_0)$ such that $u(c_1)>0$.\\
(i) if $u^{'}(c_1)>0$, then $u^{'}(y)>0$ for all $y\in(c_1,y_0)$;\\
(ii) if $u^{'}(c_1)<0$, then $u^{'}(x)<0$ for all $x\in(x_0,c_1)$. 
\end{lem}
\dem
(i) Suppose $u^{'}(c_1)>0$. Let
$$
\hat{y}=\sup\left\{y\geq c_1\:\left|\:u^{'}(z)>0,\:\:\forall z\in[c_1,y)\right.\right\}\quad.
$$
It is clear that $\hat{y}>c_1$ and
$$
u(t)\geq u(c_1)>0\quad,\quad\forall t\in[c_1,\hat{y}].
$$
From the hypothesis
$$
({\cal A}_1^V)^{*}u=\lambda u
$$
it follows that
$$
\left(\alpha u^{'}\right)^{'}=\lambda u\rho+uV\rho.
$$
Then for any $y\in(c_1,y_0)$ we have
$$
\alpha(y)u^{'}(y)-\alpha(c_1)u^{'}(c_1)=\int\limits_{c_1}^{y}\!
\rho(t)[\lambda+V(t)]u(t)\:dt\quad.
$$
If $\hat{y}<y_0$, then
$$
\alpha(\hat{y})u^{'}(\hat{y})-\alpha(c_1)u^{'}(c_1)=\int\limits_{c_1}^{\hat{y}}\!
\rho(t)[\lambda+V(t)]u(t)\:dt
$$
from where it follows that
$$
\alpha(\hat{y})u^{'}(\hat{y})=\alpha(c_1)u^{'}(c_1)+\int\limits_{c_1}^{\hat{y}}\!
\rho(t)[\lambda+V(t)]u(t)\:dt>\alpha(c_1)u^{'}(c_1)>0.
$$
Then $u^{'}(\hat{y})>0$. Hence $u^{'}(t)>0$ for all $t\in[\hat{y},\hat{y}+\varepsilon]$ for small $\varepsilon>0$, which contradicts the definition of $\hat{y}$.\\
(ii) In the same way on can prove that if $u^{'}(c_1)<0$, then $u^{'}(x)<0$, for all $x\in(x_0,c_1)$. \fin

\begin{lem}\label{23}
There exists two strictely positive functions $u_k$, $k=1,2$ on $(x_0,y_0)$ such that\\
(i) for $k=1,2$, $u_k^{'}$ is absolutely continuous and
$$
\left(\alpha u_k^{'}\right)^{'}=\lambda u_k\rho+u_kV\rho\quad\mbox{a.e.}
$$ 
where $\lambda>0$;\\
(ii) $u_1^{'}>0$ and $u_2^{'}<0$ over $(x_0,y_0)$.
\end{lem}
\dem The function $u_2$ was constructed by Feller \cite[Lemma 1.9]{feller'52} in the case where $a=1$ and $V=0$, but his prove works in the actual general framework. \fin\\ 
The main result of this subsection is
\begin{thm}\label{31}
The one-dimensional operator $({\cal A}_1^V,C_0^\infty(x_0,y_0))$ is $L^\infty(x_0,y_0;\rho dx)$-unique with respect to the topology ${\cal C}(L^\infty(x_0,y_0;\rho dx),L^1(x_0,y_0;\rho dx))$ if an only if both
$$
(*)\quad\quad\int\limits_c^{y_0}\!\rho(y)\sum\limits_{n=0}^\infty\phi_n(y)\:dy=+\infty
$$
and
$$
(**)\quad\quad\int\limits_{x_0}^c\!\rho(x)\sum\limits_{n=0}^\infty\psi_n(x)\:dx=+\infty
$$
hold, where $c\in(x_0,y_0)$, $\lambda>0$ and 
$$
\phi_n(y)=\int\limits_{c}^{y}\!\frac{1}{\alpha(r_n)}\:dr_n\int\limits_{c}^{r_n}\!\rho(t_n)[\lambda+V(t_n)]\phi_{n-1}(t_n)\:dt_n,\quad n\geq 1,\quad\phi_0(y)=1
$$
and
$$
\psi_n(x)=\int\limits_{x}^{c}\!\frac{1}{\alpha(r_n)}\:dr_n\int\limits_{r_n}^{c}\!\rho(t_n)[\lambda+V(t_n)]\psi_{n-1}(t_n)\:dt_n,\quad n\geq 1,\quad\psi_0(x)=1.
$$
\end{thm}
\dem
$\Rightarrow$ Let $({\cal A}_1^V,C_0^\infty(x_0,y_0))$ be $L^\infty(x_0,y_0;\rho dx)$-unique with respect to the topology ${\cal C}(L^\infty(x_0,y_0;\rho dx),L^1(x_0,y_0;\rho dx))$ and assume that (**) (similar in the case (*)) doesn't hold, that is
$$
\int\limits_{x_0}^c\!\rho(x)\sum\limits_{n=0}^\infty\psi_n(x)\:dx<+\infty
$$
where $c\in(x_0,y_0)$ is fixed and $\lambda>0$. We prove that there exists $u\in L^1(x_0,y_0;\rho dx)$, $u\neq 0$ such that
$$
\left[\lambda I-({\cal A}_1^V)^{*}\right]u=0\quad\mbox{\it in the sense of distributions}
$$
which is in contradiction with the $L^\infty(x_0,y_0;\rho dx)$-uniqueness of  $({\cal A}_1^V,C_0^\infty(x_0,y_0))$.\\
Indeed, by Lemma \ref{23} there exists a function $u$ strictely positive on $(x_0,y_0)$ such that $u^{'}$ is absolutely continuous, $u^{'}<0$ over $(x_0,y_0)$ and
$$
\left(\alpha u^{'}\right)^{'}=\rho(\lambda+V)u.
$$
Below we shall prove that $u\in L^1(x_0,y_0;\rho dx)$.\\
({\bf I}) {\it integrability near $y_0$}\\
For $y\in(c,y_0)$ we have
$$
\alpha(y)u^{'}(y)-\alpha(c)u^{'}(c)=\int\limits_c^y\!\rho(t)[\lambda+V(t)]u(t)\:dt.
$$
Then
$$
0\geq\alpha(y)u^{'}(y)=\alpha(c)u^{'}(c)+\int\limits_c^y\!\rho(t)[\lambda+V(t)]u(t)\:dt
$$
which implies that
$$
\int\limits_c^y\!u(t)\rho(t)\:dt\leq\int\limits_c^y\!\rho(t)[\lambda+V(t)]u(t)\:dt\leq-\alpha(c)u^{'}(c)<+\infty.
$$
({\bf II}) {\it integrability near $x_0$}\\
For $x\in(x_0,c)$ we have
$$
\alpha(c)u^{'}(c)-\alpha(x)u^{'}(x)=\int\limits_x^c\!\rho(t)[\lambda+V(t)]u(t)\:dt
$$
so that
$$
\alpha(x)u^{'}(x)=\alpha(c)u^{'}(c)-\int\limits_x^c\!\rho(t)[\lambda+V(t)]u(t)\:dt.
$$
Moreover for $c_0\in(x,c)$ we have:
$$
u(x)=u(c)-\int\limits_x^c\!u^{'}(r)\:dr=\\
$$
$$
=u(c)-\int\limits_x^c\!\left\{\frac{\alpha(c)u^{'}(c)}{\alpha(r)}-
\frac{1}{\alpha(r)}\int\limits_r^c\!\rho(t)[\lambda+V(t)]u(t)\:dt\right\}\:dr=\\
$$
$$
=u(c)-\alpha(c)u^{'}(c)\int\limits_x^c\!\frac{1}{\alpha(r)}\:dr+
\int\limits_x^c\frac{1}{\alpha(r)}\:dr\int\limits_r^c\!\rho(t)[\lambda+V(t)]u(t)\:dt=
$$
$$
=u(c)-\alpha(c)u^{'}(c)\left[\int\limits_x^{c_0}\frac{1}{\alpha(r)}\:dr+\int\limits_{c_0}^{c}\frac{1}{\alpha(r)}\:dr\right]+
$$
$$
+\int\limits_x^c\frac{1}{\alpha(r)}\:dr\int\limits_r^c\!\rho(t)[\lambda+V(t)]u(t)\:dt=
$$
$$
=u(c)-\alpha(c)u^{'}(c)\int\limits_x^{c_0}\frac{1}{\alpha(r)}\cdot
\frac{\int\limits_{c_0}^c\!\rho(t)[\lambda+V(t)]\:dt}{\int\limits_{c_0}^c\!\rho(t)[\lambda+V(t)]\:dt}\:dr-
$$
$$
-\alpha(c)u^{'}(c)\int\limits_{c_0}^{c}\frac{1}{\alpha(r)}\:dr+
\int\limits_x^c\frac{1}{\alpha(r)}\:dr\int\limits_r^c\!\rho(t)[\lambda+V(t)]u(t)\:dt=
$$
$$
=u(c)-\frac{\alpha(c)u^{'}(c)}{\int\limits_{c_0}^c\!\rho(t)[\lambda+V(t)]\:dt}\int\limits_x^{c_0}\frac{1}{\alpha(r)}\:dr
\int\limits_{c_0}^c\!\rho(t)[\lambda+V(t)]\:dt-
$$
$$
-\alpha(c)u^{'}(c)\int\limits_{c_0}^{c}\frac{1}{\alpha(r)}\:dr+
\int\limits_x^c\frac{1}{\alpha(r)}\:dr\int\limits_r^c\!\rho(t)[\lambda+V(t)]u(t)\:dt\leq
$$
$$
\leq u(c)-\frac{\alpha(c)u^{'}(c)}{\int\limits_{c_0}^c\!\rho(t)[\lambda+V(t)]\:dt}\int\limits_x^{c_0}\frac{1}{\alpha(r)}\:dr
\int\limits_{r}^c\!\rho(t)[\lambda+V(t)]\:dt- 
$$
$$
-\alpha(c)u^{'}(c)\int\limits_{c_0}^{c}\frac{1}{\alpha(r)}\:dr+
\int\limits_x^c\frac{1}{\alpha(r)}\:dr\int\limits_r^c\!\rho(t)[\lambda+V(t)]u(t)\:dt\leq
$$
$$
\leq u(c)-\frac{\alpha(c)u^{'}(c)}{\int\limits_{c_0}^c\!\rho(t)[\lambda+V(t)]\:dt}\int\limits_x^{c}\frac{1}{\alpha(r)}\:dr
\int\limits_{r}^c\!\rho(t)[\lambda+V(t)]\:dt- 
$$
$$
-\alpha(c)u^{'}(c)\int\limits_{c_0}^{c}\frac{1}{\alpha(r)}\:dr+
\int\limits_x^c\frac{1}{\alpha(r)}\:dr\int\limits_r^c\!\rho(t)[\lambda+V(t)]u(t)\:dt.
$$
Thus:
$$
u(x)\leq u(c)-\alpha(c)u^{'}(c)\int\limits_{c_0}^{c}\frac{1}{\alpha(r)}\:dr- 
$$
$$
-\frac{\alpha(c)u^{'}(c)}{\int\limits_{c_0}^c\!\rho(t)[\lambda+V(t)]\:dt}\int\limits_x^{c}\frac{1}{\alpha(r)}\:dr
\int\limits_{r}^c\!\rho(t)[\lambda+V(t)]\:dt+
$$
$$
+\int\limits_x^c\frac{1}{\alpha(r)}\:dr\int\limits_r^c\!\rho(t)[\lambda+V(t)]u(t)\:dt.
$$
If we denote
$$
M=u(c)-\alpha(c)u^{'}(c)\int\limits_{c_0}^{c}\frac{1}{\alpha(r)}\:dr, 
$$
$$
N=-\frac{\alpha(c)u^{'}(c)}{\int\limits_{c_0}^c\!\rho(t)[\lambda+V(t)]\:dt}
$$
and
$$
\psi_n(x)=\int\limits_{x}^{c}\!\frac{1}{\alpha(r_n)}\:dr_n\int\limits_{r_n}^{c}\!\rho(t_n)[\lambda+V(t_n)]\psi_{n-1}(t_n)\:dt_n,\quad n\geq 1,\quad\psi_0(x)=1
$$
then
$$
u(x)\leq M+N\psi_1(x)+\int\limits_x^c\frac{1}{\alpha(r_1)}\:dr_1\int\limits_{r_1}^c\!\rho(t_1)[\lambda+V(t_1)]u(t_1)\:dt_1.
$$
But
$$
u(t_1)\leq M+N\psi_1(t_1)+
\int\limits_{t_1}^c\frac{1}{\alpha(r_2)}\:dr_2\int\limits_{r_2}^c\!\rho(t_2)[\lambda+V(t_2)]u(t_2)\:dt_2.
$$
By iteration we obtain:
$$
u(x)\leq M+N\psi_1(x)+
M\int\limits_x^c\frac{1}{\alpha(r_1)}\:dr_1\int\limits_{r_1}^c\!\rho(t_1)[\lambda+V(t_1)]\:dt_1+
$$
$$
+N\int\limits_x^c\frac{1}{\alpha(r_1)}\:dr_1\int\limits_{r_1}^c\!\rho(t_1)[\lambda+V(t_1)]\psi_1(t_1)\:dt_1+
$$
$$
+\int\limits_x^c\frac{1}{\alpha(r_1)}\:dr_1\int\limits_{r_1}^c\!\rho(t_1)[\lambda+V(t_1)]\:dt_1
\int\limits_{t_1}^c\frac{1}{\alpha(r_2)}\:dr_2\int\limits_{r_2}^c\!\rho(t_2)[\lambda+V(t_2)]u(t_2)\:dt_2\leq
$$
$$
\leq(M+N)\psi_0(x)+(M+N)\psi_1(x)+N\psi_2(x)+
$$
$$
+\int\limits_x^c\frac{1}{\alpha(r_1)}\:dr_1\int\limits_{r_1}^c\!\rho(t_1)[\lambda+V(t_1)]\:dt_1
\int\limits_{t_1}^c\frac{1}{\alpha(r_2)}\:dr_2\int\limits_{r_2}^c\!\rho(t_2)[\lambda+V(t_2)]u(t_2)\:dt_2\leq\cdots
$$
$$
\cdots\leq(M+N)\sum\limits_{n=0}^\infty\psi_n(x).
$$
Hence
$$
\int\limits_{x_0}^c\!u(x)\rho(x)\:dx\leq(M+N)\int\limits_{x_0}^c\!\rho(x)\sum\limits_{n=0}^\infty\psi_n(x)\:dx<+\infty.
$$
This show the $\rho$-integrability of $u$ near $x_0$.\\
$\Leftarrow$ Assume that (*) and (**) hold. Suppose in contrary that $({\cal A}_1^V,C_0^\infty(x_0,y_0))$ is not $L^\infty(x_0,y_0;\rho dx)$-unique. Then there exists
$h\in L^1(x_0,y_0;\rho dx)$, $h\neq 0$ which satisfies
$$
\left(\lambda I-({\cal A}_1^V)^{*}\right)h=0
$$
for some $\lambda>0$. We can assume that $h\in C^1(x_0,y_0)$ and $h>0$ on some interval $[x_1,y_1]\subset(x_0,y_0)$, where $x_1<y_1$. Notice that $h^{'}\neq 0$ on $(x_1,y_1)$.\\
Let $c_1\in(x_1,y_1)$.\\
({\bf I}) {\it case} $h^{'}(c_1)>0$.\\ 
By Lemma \ref{22}, it follows
$$
h^{'}(y)>0\quad,\quad\forall y\in(c_1,y_1).
$$
Hence
$$
h(y)\geq h(c_1)>0\quad,\quad\forall y\in[c_1,y_1].
$$
Then we have:
$$
h(y)=h(c_1)+\int\limits_{c_1}^y\!h^{'}(r)\:dr=\\
$$
$$
=h(c_1)+\int\limits_{c_1}^y\!\left\{\frac{\alpha(c_1)h^{'}(c_1)}{\alpha(r)}+\frac{1}{\alpha(r)}
\int\limits_{c_1}^r\!\rho(t)[\lambda+V(t)]h(t)\:dt\right\}\:dr>\\
$$
$$
>h(c_1)+\int\limits_{c_1}^y\!\frac{1}{\alpha(r)}\:dr\int\limits_{c_1}^r\!\rho(t)[\lambda+V(t)]h(t)\:dt.
$$
Using inductively this inequality we get
$$
h(y)>h(c_1)+\int\limits_{c_1}^y\!\frac{1}{\alpha(r_1)}\:dr_1\int\limits_{c_1}^{r_1}\!\rho(t_1)[\lambda+V(t_1)]h(t_1)\:dt_1>
$$
$$
>h(c_1)+h(c_1)\int\limits_{c_1}^y\!\frac{1}{\alpha(r_1)}\:dr_1\int\limits_{c_1}^{r_1}\!\rho(t_1)[\lambda+V(t_1)]\:dt_1+
$$
$$
+\int\limits_{c_1}^y\!\frac{1}{\alpha(r_1)}\:dr_1\int\limits_{c_1}^{r_1}\!\rho(t_1)[\lambda+V(t_1)]\:dt_1
\int\limits_{c_1}^{t_1}\!\frac{1}{\alpha(r_2)}\:dr_2\int\limits_{c_1}^{r_2}\!\rho(t_2)[\lambda+V(t_2)]h(t_2)\:dt_2>\cdots
$$
$$
\cdots>h(c_1)\sum\limits_{n=0}^{\infty}\phi_n(y).
$$
Consequently
$$
\int\limits_{x_0}^{y_0}\!h(y)\rho(y)\:dy\geq\int\limits_{c_1}^{y_0}\!h(y)\rho(y)\:dy >h(c_1)\int\limits_{c_1}^{y_0}\!\rho(y)\sum\limits_{n=0}^{\infty}\phi_n(y)\:dy=+\infty
$$
which is a contradiction with the assumption $h\in L^1(x_0,y_0;\rho dx)$.\\
({\bf II}) {\it case} $h^{'}(c_1)<0$.\\
We prove in a similar way that
$$
\int\limits_{x_0}^{y_0}\!h(x)\rho(x)\:dx>+\infty.\quad \square
$$
In particular, for $V=0$, the one-dimensional operator
$$
{\cal A}_1f=a(x)f^{''}+b(x)f^{'}
$$
is $L^\infty(x_0,y_0;\rho dx)$-unique with respect to the topology ${\cal C}(L^\infty(x_0,y_0;\rho dx),L^1(x_0,y_0;\rho dx))$ if an only if both
$$
(\circ)\quad\quad\int\limits_c^{y_0}\!\rho(y)\:dy\int\limits_{c}^{y}\!\frac{1}{\alpha(r)}\:dr
\int\limits_{c}^{r}\!\rho(t)\:dt=+\infty
$$
and
$$
(\circ\circ)\quad\quad\int\limits_{x_0}^c\!\rho(x)\:dx\int\limits_{x}^{c}\!\frac{1}{\alpha(r)}\:dr
\int\limits_{r}^{c}\!\rho(t)\:dt=+\infty
$$
hold. In the terminology of Feller this means that $y_0$ and, respectively $x_0$ are {\it no entrance boundaries} (see \cite[Theorem 4.1,p.590]{wu-zhang'06}).

\subsection{The multidimensional case}
In this subsection we consider the multidimensional generalized Schr\"odinger operator
$$
{\cal A}^Vf:=\frac{1}{2}\Delta f+b\cdot\nabla f-Vf\quad,\quad\forall f\in C_0^\infty(\R^d)
$$
where $d\geq 2$ and $V$ is non-negative. Denote the euclidian norm in $\R^d$ by $|x|=\sqrt{x\cdot x}$. If there is some mesurable locally bounded function 
$$
\beta:\R^{+}\rightarrow\R
$$
such that
$$
b(x)\cdot\frac{x}{|x|}\geq\beta(|x|)\quad,\quad\forall x\in\R^d,\:x\neq 0,
$$
then for any initial point $x\neq 0$ we have
$$
|X_t|-|x|\geq\int\limits_0^t\!\left[\beta(|X_t|)+\frac{d-1}{2|X_t|}\right]\:dt+\mbox{\it a real Brownian motion},\quad\forall t<\tau_e.
$$
In other words, $|X_t|$ go to infinity more rapidly than the one-dimensional diffusion generated by 
$$
{\cal A}_1=\frac{1}{2}\frac{d^2}{dr^2}+\left[\beta(r)+\frac{d-1}{2r}\right]\frac{d}{dr}.
$$
This is standard in probability (see {\sc Ikeda}, {\sc Watanabe} \cite {ikeda-watanabe'81}). Remark that for the one-dimensional operator
$$
{\cal A}_1^V=\frac{1}{2}\frac{d^2}{dr^2}+\left[\beta(r)+\frac{d-1}{2r}\right]\frac{d}{dr}-V(r)
$$
the speed measure of Feller is given by
$$
\rho(r)=2e^{\int\limits_1^r\!2\left[\beta(t)+\frac{d-1}{2t}\right]\:dt}=
2e^{\int\limits_1^r\!2\beta(t)\:dt}e^{\int\limits_1^r\!\frac{d-1}{t}\:dt}=
2r^{d-1}e^{\int\limits_1^r\!2\beta(t)\:dt}
$$
and the scale function of Feller is
$$
\alpha(r)=r^{d-1}e^{\int\limits_1^r\!2\beta(t)\:dt}.
$$
Now we can formulate the main result of this subsection:  
\begin{thm}\label{31}
Suppose that there is some mesurable locally bounded function 
$$
\beta:\R^{+}\rightarrow\R
$$
such that
$$
b(x)\cdot\frac{x}{|x|}\geq\beta(|x|)\quad,\quad\forall x\in\R^d,\:x\neq 0.
$$
If the one-dimensional diffusion operator
$$
{\cal A}_1^V=\frac{1}{2}\frac{d^2}{dr^2}+\left[\beta(r)+\frac{d-1}{2r}\right]\frac{d}{dr}-V(r)
$$
is $L^\infty(0,\infty;\rho dx)$-unique with respect to the topology ${\cal C}(L^\infty(0,\infty;\rho dx),L^1(0,\infty;\rho dx))$, then the generalized Schr\"odinger operator $\left({\cal A}^V,C_0^\infty(\R^d)\right)$ is $L^\infty(\R^d,dx)$-unique with respect to the topology ${\cal C}(L^\infty,L^1)$.
\end{thm}
\dem
By Theorem \ref{11}, for the $L^\infty(\R^d,dx)$-uniqueness of $\left({\cal A}^V,C_0^\infty(\R^d)\right)$ it is enough to show that if for some $\lambda>0$, $u\in L^1(\R^d,dx)$ satisfies
$$
\left(({\cal A}^V)^{*}-\lambda I\right)u=0\quad\mbox{\it in the sense of distributions}
$$
then $u=0$.\\
Let $\lambda>0$ and $u\in L^1(\R^d,dx)$ such that
$$
\left\langle u,\left({\cal A}^V-I\right)f\right\rangle=0\quad,\quad\forall f\in C_0^\infty(\R^d)
$$
where
$$
\langle f,g\rangle:=\int\limits_{\R^d}\!fg\:dx.
$$
The above equality becomes
$$
\frac{1}{2}\int\limits_{\R^d}\!u(x)\Delta f(x)\:dx+\int\limits_{\R^d}\!u(x)b\cdot\nabla f(x)\:dx=\int\limits_{\R^d}\!u(x)(\lambda+V)f(x)\:dx=0\quad,\quad\forall f\in C_0^\infty(\R^d).
$$
By the ellipticity regularity result in \cite[Lemma 2, p.341]{eberle'00}, $u\in L_{loc}^\infty(\R^d)$ and $\nabla u\in L_{loc}^d(\R^d)\subset L_{loc}^2(\R^d)$. By the fact that $C_0^\infty(\R^d)$ is dense in
$$
\left\{\left.f\in L^2\:\right|\:\nabla f\in L^2\mbox{ and the support of }f\mbox{ is compact}\right\}
$$
an integration by parts yields
$$
-\frac{1}{2}\int\limits_{\R^d}\!\nabla u(x)\cdot\nabla f(x)\:dx+\int\limits_{\R^d}\!u(x)b\cdot\nabla f(x)\:dx=\int\limits_{\R^d}\!u(x)(\lambda+V)f(x)\:dx
$$
for all $f\in H^{1,2}(\R^d)$ with compact support. Now on can follow {\sc Eberle} \cite[proof of Theorem 1, 335]{eberle'00} to show the next inequality of Kato's type
$$
-\frac{1}{2}\int\limits_{\R^d}\!\nabla|u(x)|\cdot\nabla f(x)\:dx+\int\limits_{\R^d}\!|u(x)|b\cdot\nabla f(x)\:dx\geq\int\limits_{\R^d}\!|u(x)|(\lambda+V)f(x)\:dx
$$
for all $f\in H^{1,2}(\R^d)$ with compact support.\\
Let
$$
G(r)=\int\limits_{B(r)}\!|u(x)|\:dx
$$
where $B(r)=\left\{\left.x\in \R^d\:\right|\:|x|\leq r\right\}$. $G$ is absolutely continuous and
$$
G^{'}(r)=\int\limits_{\partial B(r)}\!|u(x)|\:d_\sigma x\quad,\quad\mbox{dr-a.e.}
$$
where $d_\sigma r$ is the surface measure on the sphere $\partial B(r)$ (the boundary of $B(r)$). Now for every $0<r_1<r_2$ we consider
$$
f=\min\left\{r_2-r_1,(r_2-|x|)^{+}\right\}
$$
and
$$
\gamma(x)=\frac{x}{|x|}=\nabla |x|\quad.
$$ 
Then we have
$$
-\frac{1}{2}\int\limits_{B(r_2)-B(r_1)}\!\nabla|u(x)|\cdot\nabla(r_2-|x|)\:dx+
\int\limits_{B(r_2)-B(r_1)}\!|u(x)|b(x)\cdot\nabla(r_2-|x|)\:dx\geq
$$
$$
\geq\int\limits_{B(r_2)-B(r_1)}\!|u(x)|(\lambda+V)(r_2-|x|)\:dx
$$
from where it follows that
$$
\frac{1}{2}\int\limits_{B(r_2)-B(r_1)}\!\nabla|u(x)|\cdot\gamma(x)\:dx-
\int\limits_{B(r_2)-B(r_1)}\!|u(x)|b(x)\cdot\gamma(x)\:dx\geq
$$
$$
\geq\int\limits_{B(r_2)-B(r_1)}\!|u(x)|(\lambda+V)(r_2-|x|)\:dx\quad.
$$
Since
$$
\nabla|u|\gamma=div(|u|\gamma)-|u|div(\gamma)=div(|u|\gamma)-|u|\frac{d-1}{|x|},
$$
by the Gauss-Green formula we have
$$
\int\limits_{B(r_2)-B(r_1)}\!\nabla|u(x)|\cdot\gamma(x)\:dx=G^{'}(r_2)-G^{'}(r_1)-
(d-1)\int\limits_{r_1}^{r_2}\!\frac{1}{r}G^{'}(r)\:dr
$$
for $dr_1\otimes dr_2$-a.e. $0<r_1<r_2$.\\
By another hand, using the hypothese
$$
b(x)\cdot\gamma(x)=b(x)\cdot\frac{x}{|x|}\geq\beta(|x|)
$$
and Fubini's theorem, we get
$$
-\int\limits_{B(r_2)-B(r_1)}\!|u(x)|b(x)\cdot\gamma(x)\:dx\leq
-\int\limits_{r_1}^{r_2}\!G^{'}(r)\beta(r)\:dr
$$
and
$$
\int\limits_{B(r_2)-B(r_1)}\!|u(x)|(\lambda+V)(r_2-|x|)\:dx=
\int\limits_{r_1}^{r_2}\![\lambda+V(r)](r_2-r)G^{'}(r)\:dr=
$$
$$
=\int\limits_{r_1}^{r_2}\![\lambda+V(r)]G^{'}(r)\int\limits_{r}^{r_2}\!dt\:dr=
\int\limits_{r_1}^{r_2}\!dr\int\limits_{r_1}^{r}\![\lambda+V(t)]G^{'}(t)\:dt.
$$
Consequently
$$
\frac{1}{2}\left[G^{'}(r_2)-G^{'}(r_1)\right]-
\int\limits_{r_1}^{r_2}\!\left[\beta(r)+\frac{d-1}{2r}\right]G^{'}(r)\:dr\geq
$$
$$
\geq\int\limits_{r_1}^{r_2}\!dr\int\limits_{r_1}^{r}\![\lambda+V(t)]G^{'}(t)\:dt
$$
for $dr_1\otimes dr_2$-a.e. $0<r_1<r_2$.\\
Consider the differential form
$$
{\cal A}_1^{-}:=\frac{1}{2}G^{''}(r)-\left[\beta(r)+\frac{d-1}{2r}\right]G^{'}(r)
$$
in the sense of distribution on $(0,\infty)$. Notice that the sign of $\beta(r)+\frac{d-1}{2r}$ in ${\cal A}^{-}_1$ is negative, opposite to the sign in the operator ${\cal A}^V_1$ and the speed measure of Feller for ${\cal A}^{-}_1$ is exactely $\rho(r)$ and the scale function of Feller for ${\cal A}^{-}_1$ is $\alpha(r)$. Hence we can write ${\cal A}^{-}_1$ in the Feller form
$$
{\cal A}^{-}_1=\frac{1}{2}G^{''}-\left[\beta(r)+\frac{d-1}{2r}\right]G^{'}=\frac{1}{2}G^{''}-\frac{\alpha^{'}}{\rho}G^{'}=
$$
$$
=\frac{1}{2}G^{''}-\frac{\rho^{'}}{2\rho}G^{'}=\frac{\rho}{2}\frac{\rho G^{''}-\rho^{'}G^{'}}{\rho^2}=\alpha\left(\frac{G^{'}}{\rho}\right)^{'}.
$$
Then we have
$$
\left(\frac{G^{'}}{\rho}\right)^{'}\geq\frac{1}{\alpha}\int\limits_{r_1}^{r_2}\![\lambda+V(t)]G^{'}(t)\:dt
$$
in the sense of distribution on $(0,\infty)$.\\
Assume now in contrary that $u\neq 0$. Then there exists $c\in(r_1,r_2)$ such that $G^{'}(c)>0$. Then for $dy$-a.e. $y>c$ we have
$$
\frac{G^{'}}{\rho}(y)\geq\frac{G^{'}}{\rho}(c)+\int\limits_{c}^{y}\!\frac{1}{\alpha(r)}\:dr
\int\limits_{c}^{r}\![\lambda+V(t)]G^{'}(t)\:dt=
$$
$$
=\frac{G^{'}}{\rho}(c)+\int\limits_{c}^{y}\!\frac{1}{\alpha(r)}\:dr
\int\limits_{c}^{r}\!\rho(t)[\lambda+V(t)]\frac{G^{'}}{\rho}(t)\:dt.
$$
Using the above inequality inductively we get
$$
\frac{C^{'}}{\rho}(y)\geq\frac{G^{'}}{\rho}(c)\sum\limits_{n=0}^{\infty}\phi_n(y)
$$
where $\phi_0(y)=1$ and for any $n\in\N^{*}$,
$$
\phi_n(y)=\int\limits_{c}^{y}\!\frac{1}{\alpha(r_n)}\:dr_n\int\limits_{c}^{r_n}\!\rho(t_n)[\lambda+V(t_n)]\phi_{n-1}(t_n)\:dt_n.
$$
By Theorem \ref{31} it follows that
$$
\int\limits_{\R^d}\!|u(x)|\:dx=G(\infty)\geq\frac{G^{'}}{\rho}(c)\int\limits_{c}^{\infty}\!\rho(y)\sum\limits_{n=0}^{\infty}\phi_n(y)\:dy=+\infty
$$
because ${\cal A}^V_1$ is suppose to be $L^\infty(0,\infty;\rho dx)$-unique. This in contradiction with the assumption that $u\in L^1(\R^d,dx)$. \fin \\ 
Remark that if $\cal A$ is a second order elliptic differential operator with ${\cal D}=C_0^\infty(\R^d)$, then the weak solutions for
the dual Cauchy problem in the Theorem \ref{11} (v) correspond
exactly to those in the distribution sense in the theory of partial
differential equations and the dual Cauchy problem becomes the Fokker-Planck equation for heat diffusion. Then we can formulate
\begin{cor}
In the hypothesis of Theorem \ref{31}, for any $f\in L^1(\R^d,dx)$ the Fokker-Planck equation
$$
\left\lbrace\begin{array}{l}
\partial_tu(t,x)=\frac{1}{2}\Delta u(t,x)-div\left(bu(t,x)\right)-Vu(t,x)\\
u(0,x)=f(x)
\end{array}
\right.
$$
has one $L^1(\R^d,dx)$-unique weak solution.
\end{cor} 
\dem The assertion follows by the Theorem \ref{11} and the Theorem
\ref{31}. \fin

\bibliographystyle{plain}

\begin{flushright}
Engineering Faculty of Hunedoara,\\
"Politehnica" University of Timi\c soara,\\
331128 Hunedoara, Romania\\
and\\ 
Institut Camille Jordan (CNRS-UMR 5208),\\
Universit\'e Claude Bernard Lyon1,\\
69622 Villeurbanne, France.\\
\end{flushright}

\begin{flushright}
e-mail: {\tt lemle.dan@fih.upt.ro}
\end{flushright}

\end{document}